\documentclass[lettersize,journal]{IEEEtran}
\usepackage{amsmath,amsfonts}
\usepackage{algorithmic}
\usepackage{algorithm}
\usepackage{array}
\usepackage{stfloats}
\usepackage{url}
\usepackage{graphicx}
\usepackage{cite}
\usepackage{balance}
\usepackage{hyperref}
\hypersetup{colorlinks=true, breaklinks=true, linkcolor=black, citecolor=black, urlcolor=blue} 
\usepackage{braket}
\usepackage{multirow}
\usepackage{booktabs}
\usepackage{colortbl}
\usepackage{ragged2e}
\usepackage{subcaption}
\usepackage{caption}
\usepackage{balance}
\usepackage{makecell}

\begin{document}


\title{GraphDART: Graph Distillation for Efficient Advanced Persistent Threat Detection}

\author{
Saba Fathi Rabooki, Bowen Li, Falih Gozi Febrinanto, Ciyuan Peng,~\IEEEmembership{Student Member,~IEEE}, Elham Naghizade, Fengling Han,~\IEEEmembership{Senior Member,~IEEE}, and Feng Xia,~\IEEEmembership{Senior Member,~IEEE}

\thanks{S. F. Rabooki, B. Li, E. Naghizade, F. Han, and F. Xia are with School of Computing Technologies, RMIT University, Melbourne, VIC 3000, Australia. (e-mail: saba.fathi.rabooki@student.rmit.edu.au, \{bowen.li; e.naghizade; fengling.han\}@rmit.edu.au, f.xia@ieee.org)}
\thanks{F. G. Febrinanto and C. Peng are with Institute of Innovation, Science and Sustainability, Federation University Australia, Ballarat, VIC 3353, Australia. (e-mail: f.febrinanto@federation.edu.au, ciyuan.p@ieee.org)}
}

\markboth{
}%
{Fathi \MakeLowercase{\textit{et al.}}: GraphDART: Graph Distillation for Efficient Advanced Persistent Threat Detection}

\maketitle





\begin{abstract}

Cyber-physical-social systems (CPSSs) have emerged in many applications over recent decades, requiring increased attention to security concerns. The rise of sophisticated threats like Advanced Persistent Threats (APTs) makes ensuring security in CPSSs particularly challenging. Provenance graph analysis has proven effective for tracing and detecting anomalies within systems, but the sheer size and complexity of these graphs hinder the efficiency of existing methods, especially those relying on graph neural networks (GNNs). To address these challenges, we present GraphDART, a modular framework designed to distill provenance graphs into compact yet informative representations, enabling scalable and effective anomaly detection. GraphDART can take advantage of diverse graph distillation techniques, including classic and modern graph distillation methods, to condense large provenance graphs while preserving essential structural and contextual information. This approach significantly reduces computational overhead, allowing GNNs to learn from distilled graphs efficiently and enhance detection performance. Extensive evaluations on benchmark datasets demonstrate the robustness of GraphDART in detecting malicious activities across cyber-physical-social systems. By optimizing computational efficiency, GraphDART provides a scalable and practical solution to safeguard interconnected environments against APTs.

\end{abstract}


\section{Introduction}

\IEEEPARstart{C}{yber}-physical-social systems (CPSSs) have become increasingly intertwined in various aspects of human life, with applications in diverse areas, including, e.g., transportation, smart manufacturing, sustainability, energy distribution, and healthcare \cite{cpss_2023_complexHumanMachineSystem, CPSS_2018,XiaBeeInfo2015}. 
However, due to the complex nature of CPSSs, which span multiple domains from physical processes to computational space and human interaction, security issues with severe consequences are prevalent~\cite{CPSS_2020, NajaflouSafety2015}. 
For instance, a security breach at Maroochy Water led to the release of 800,000 liters of raw sewage into the local community and natural waterways, causing significant environmental damage and subsequent economic costs~\cite{MaroochyWaterAttack_2008}.

One of the security vulnerabilities that are of particular concern in CPSS is Advanced Persistent Threats (APTs). 
APTs are prolonged cyberattacks designed to stealthily exfiltrate sensitive information from target systems or disrupt critical functions~\cite{APT_2019}. 
A notable example is the 2015 Ukraine blackout~\cite{APTExample_2017}, where APT attackers broke into the power control system through spear-phishing (targeting the social domain of a CPSS) and executed multiple cyberattacks to compromise critical power infrastructure (threatening cyber and physical domains)~\cite{CPSS_APT_2024}. 
APT campaigns target the entire systems rather than individual components, employing advanced techniques over prolonged periods to orchestrate and exploit sophisticated multi-stage attacks~\cite{APT_2019}. 
Since CPSSs integrate three dimensions (namely, cyber, physical, and social), they are exposed to vulnerabilities across all domains.

Whilst the study of security is well-established in cyber-physical and cyber-social systems, the complex nature of APTs may lead to the failure of traditional Intrusion Detection Systems (IDSs) to accurately detect such threats, especially in a complex CPSS. 
APT attackers employ a low-and-slow strategy, mimicking the normal behaviour of the target system during attack stages to evade detection, which over long time, has been described as ``\textit{finding a needle in a haystack}"~\cite{sok_2023_prvGraph}. This presents a significant challenge for traditional security defense mechanisms.

Such low-and-slow strategy requires modeling fine-grained, longitudinal activities, such as raw system logs, in a way to better capture the underlying causal relationships between \textit{entities}, i.e., processes, files, or network sockets, and \textit{events}, i.e., activities between entities. 
However, since auditing systems capture events sequentially ordered by timestamp, raw system logs typically lack causality relationships among system entities, making it difficult to trace the flow of actions~\cite{sok_2023_prvGraph}.

In contrast, a provenance graph captures these causal relationships among system entities. In such a graph, entities and events are mapped to the nodes and edges of the graph. The edges are directed and represent the flow of information among nodes (more details can be found in Section~\ref{sec:PrvGraph}).

Despite the advantage of provenance graph representation, processing such graph becomes a challenge when its size grows significantly \cite{FebrinantoGLL2023}. The persistence of APT requires keeping track of the entire history of system events, which leads to a large provenance graph. For example, a single log file \cite{dataset_darpa3} with 5 million records capturing events in just three hours can correspond to a provenance graph with approximately 300 thousand nodes and 9 million edges. 

Since APTs can remain undetected in a target system for over months, connecting the dots to track the progression of APT requires processing an exponentially growing volume of data as more information accumulates over time. 
This increasing data volume imposes higher computational costs and challenges the effectiveness of the detection tools.

APT detection tools can fall into three categories, namely, anomaly detection (e.g. \cite{AptDtc_unicorn_2020, AptDtc_atlas_2021}), rule-based (e.g. \cite{AptDtc_holmes_2019}), and scoring (e.g. \cite{AptDtc_priotracker_2018, AptDtc_morse_2020, AptDtc_depimpact_2022}). 
For instance, Unicorn \cite{AptDtc_unicorn_2020} converts the provenance graph into histograms and applies clustering to detect deviation from normal behavior in the system over time. 
Holmes \cite{AptDtc_holmes_2019} lifts up the provenance graph into a high-level APT scenario leveraging pre-defined attack signatures. 
Priotracker \cite{AptDtc_priotracker_2018} computes priority scores for events based on some factors such as rareness and degree of nodes. 
Since all these APT detection methods process very large provenance graphs, there will be computation and memory overhead in detection phase. To mitigate this runtime cost, leveraging a smaller graph with comparable detection performance to the originally huge provenance graph can be beneficial.

Therefore, provenance graph pruning techniques have been proposed to reduce the size of the graph~\cite{sok_2023_prvGraph}. 
NodeMerge \cite{GrphReduct_nodemerge_2018} considers globally read-only files (e.g. shared object libraries) as redundant in the provenance graph and aggregates set of files that are frequently accessed together. The approach proposed in \cite{GrphReduct_iburst_2016} eliminates bursty nodes such as processes that make intense bursts of events in a small time window. Hossain et al.  \cite{GrphReduct_dppreserve_2018} proposed graph reduction methods including aggregation of parallel edges. 
However, the number of nodes and edges that can be reduced by these methods are limited. Furthermore, the effectiveness of these methods are highly dependent on the features of the original provenance graph (e.g. proportion of read-only files and ratio of parallel flows between nodes.

To address these challenges, we propose \textbf{GraphDART}, a \textbf{Graph} \textbf{D}istillation-based framework for \textbf{A}dvanced pe\textbf{R}sistent \textbf{T}hreat detection. 
Graph distillation empowers GraphDART to reduce the size of provenance graphs considerably, independent from the graph characteristics, projecting its scalability. 
Through extensive experiments, we show the distilled graph preserve necessary knowledge for APT detection by producing comparable results with other state-of-the-art baselines.

The contributions of this paper are summarized as follows:
\begin{itemize}
    \item We introduced \textit{a new graph distillation framework} for the APT detection task, namely GraphDART. This framework significantly reduces the size of provenance graphs generated from log files and addresses the high computational costs of processing large-scale graphs. To our knowledge, this is the first use of graph distillation for APT detection.
    \item We introduce \textit{benign-only training strategy} to perform graph distillation using the original large-scale graph consisting exclusively of benign samples without attacks. This process generates a distilled, synthesized graph with free attack nodes, which is then used for effective and efficient APT detection training on the smaller graph.
    \item Our extensive experiments demonstrate that \textit{the proposed framework outperforms state-of-the-art techniques for APT detection}, achieving better or comparable performance while using a distilled graph that is only 5\% or less compared to the original provenance graph size.
\end{itemize}

The rest of the paper is organized as follows: First, we review the related work in Section~\ref{sec:related_work} and explain the necessary concepts and notations in the preliminaries of Section~\ref{sec:peliminaries}. After defining the problem statement in Section~\ref{sec:problem_stmnt}, we present the design of our framework in Section~\ref{sec:proposed_framework} and its evaluation in Section~\ref{sec:evaluation}. Finally, Section~\ref{sec:conclusion} concludes the paper.

\section{Related Work}
\label{sec:related_work}

\subsection{Advanced Persistent Threat Detection}

APTs present significant security challenges to organizations due to their sophisticated and persistent nature~\cite{APT_2019,sharma2023advanced}. In particular, APTs continue to challenge network security by leveraging a multi-stage attack approach and exploiting unknown vulnerabilities within target networks. To address these challenges, many machine learning and deep learning-based cybersecurity methodologies are used to detect and monitor APTs~\cite{APT_2019}. These methodologies leverage techniques such as anomaly detection, behavioral analysis, and predictive modeling to identify and respond to sophisticated attack patterns effectively. For example, Joloudari et al.~\cite{joloudari2020early} proposed a 6-layer deep learning model for early APT detection. Moreover, Lad~\cite{lad2024harnessing} explored the integration of machine learning techniques into cybersecurity for APT detection. Methods such as supervised learning and reinforcement learning were introduced to identify and mitigate APTs.

Although typical machine learning-based APT detection methods have demonstrated great potential, they face significant challenges due to the complexity of the data, such as the intricate interactions between entities and their dynamic relationships~\cite{AptDtc_Magic_2024}. To address these challenges, many graph learning-based APT detection methods have been proposed, leveraging the ability of graph-based models to capture these complex structures effectively. MAGIC~\cite{AptDtc_Magic_2024} utilized self-supervised graph learning method for detecting APTs. MAGIC constructed provenance graphs from system logs and employed graph embedding and outlier detection techniques. RAPTOR~\cite{kumar2023raptor} detected APTs in industrial internet of things (IIoT) environments. It constructed a high-level APT campaign graph to aid cybersecurity analysts in understanding and mitigating attack progressions. FLASH~\cite{flash_2024}, a scalable intrusion detection system, leveraged graph neural networks (GNNs) to detect APTs from provenance graphs. FLASH employed Word2Vec~\cite{word2vec_model_2013} for semantic encoding and GNNs for capturing structural and contextual features, achieving real-time APT detection through embedding recycling and selective graph traversal techniques. KAIROS~\cite{AptDtc_Kairos_2024} leveraged GNNs to detect and reconstruct APTs in real-time. It addressed challenges in scope, attack agnosticity, timeliness, and attack reconstruction by generating compact summary graphs from provenance data.

Due to the large-scale nature of provenance graph data, current methods often struggle with low efficiency in APT detection. To overcome this challenge, this paper proposes a novel graph distillation method to reduce data size and enable scalable learning, thereby achieving efficient APT detection.

\subsection{Graph Distillation}
 
Graph condensation aims to generate a smaller graph, allowing model training or inference on the condensed graph while maintaining performance comparable to that on the original graph~\cite{gd_GCond_2022}. For instance, GCDM~\cite{gd_gcdm_2022} is a graph condensation method based on receptive field distribution matching. It effectively reduces the size of large graphs while preserving essential structural information, thereby improving learning efficiency and maintaining model performance.

More recently, inspired by knowledge distillation~\cite{gou2021knowledge,hao2024one}, which transfers knowledge from a large model to a lightweight one, graph distillation~\cite{feng2023fair,fan2024dgsd} has emerged as a critical approach to improving the efficiency and scalability of graph learning \cite{XiaGL2021,FebrinantoGLL2023}. For instance, Kelvinius et al.~\cite{ekstrom2024accelerating} utilized graph distillation to accelerate molecular GNNs for energy and force prediction tasks. They proposed feature-based distillation strategies, such as node-to-node, edge-to-edge, and vector-to-vector distillation, enhancing the performance of lightweight student models while maintaining high inference throughput. SGDD~\cite{GC_SGDD_2024} is a graph distillation framework that broadcasts original graph structure information to synthetic condensed graphs. By minimizing Laplacian energy distribution (LED) shifts using optimal transport, SGDD supports various graph-based tasks, including node classification, anomaly detection, and link prediction, while reducing graph size. Particularly, graph distillation has been widely used in anomaly detection tasks. For example, RG-GLD~\cite{zhou2024reconstructed}, a GNN-based anomaly detection framework, integrates with graph distillation for lightweight IoT network security. It combines local subgraph preservation and global information alignment mechanisms to enhance feature representation, achieving reduced computational load and effective anomaly detection tasks.

This paper explores graph distillation to enhance the computational efficiency and scalability of APT detection in large-scale environments. We propose a new graph distillation method to reduce the size of provenance graphs while preserving essential information, such as structure, contextual information, and node distribution.

\section{Preliminaries}
\label{sec:peliminaries}

This section focuses on explaining the concepts and definitions for this work. First, system logs are introduced which provide rich information about event history of a target system. Having logs files, we can capture causal relationships among the events into a graph representation called provenance graph. Since system logs and provenance graph for detecting a long-term threat such as APT are huge and complex, graph distillation can be utilized to mitigate the storage and processing costs.

\begin{figure*}[!htbp]
    \centerline{\includegraphics[width=\linewidth]{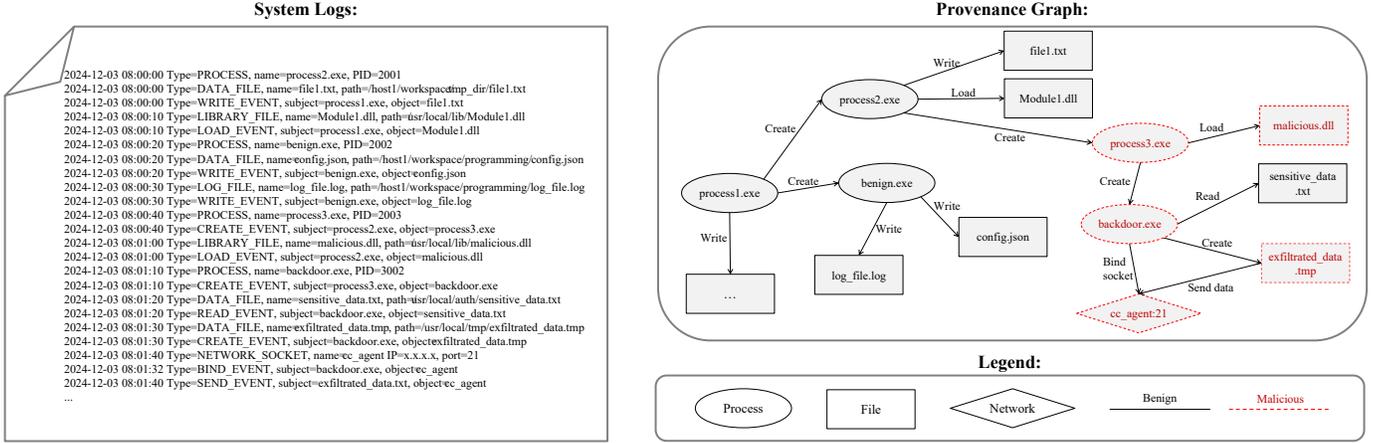}}
    \caption{Example of system logs and corresponding provenance graph. Entities and events captured in the logs are represented as nodes and edges, respectively, in the provenance graph. Differentiating between benign and malicious nodes necessitates using a threat detection tool.}
    \label{fig:pg_example}
\end{figure*}

\subsection{System Logs}

System logs record the history of events in a system and help APT detection tools to track security violations in the system \cite{sok_2023_prvGraph}. 
System logs contain data records describing \textbf{entities} and \textbf{events}, explained as follows \cite{prvGraphDtc_survey_2021}:
\begin{itemize}
    \item \textbf{Entities} (Subjects and Objects): A \textbf{subject} is an entity that performs an action on an \textbf{object}. An object in one operation can serve as a subject in another event. For instance, when a process writes to a file, the process is the subject performing the `write' action on the file, which is the object.
    
    \item \textbf{Events}: The operation that a subject performs on an object is called an \textbf{event}. Event $event_i$ can be represented as follows:
        \begin{equation}
            event_i = \braket{u_i, v_i, t_i, o_i}
            \label{EventDefinitioin}
        \end{equation}
    where $u_i$, $v_i$, $t_i$, and $o_i$ represent the subject, object, timestamp, and operation, respectively. $|E|$ denotes the total number of events in the log file, $i \in [0, |E|-1]$.
\end{itemize}

\subsection{Provenance Graph}
\label{sec:PrvGraph}

Provenance graph represents causal relation of events in form of a directed graph~{\cite{prvIdsSurvey_2022}}. Entities in system logs are mapped to \textbf{nodes} in provenance graph, and events become \textbf{edges} of the graph directed from corresponding subject to object. 
Each node and edge contain a set of attributes, represented as follows:

\begin{equation}
    n_i = \braket{nid_i, ntype_i, nattr_i^*}
    \label{NodeDefinition}
\end{equation}
where each node $n$ has a unique ID $nid$, a type $ntype \in \Set{\text{Types of system entities}}$, and other optional attributes $nattr^*$ depending on the data provided by each dataset or auditing tool in a target system. 

\begin{equation}
    e_i = \braket{eid_i, etype_i, srcnid_i, dstnid_i, t_i, eattr_i^*}
    \label{EdgeDefinition}
\end{equation}
where each edge $e$ has a unique ID $eid$, a type $etype \in \Set{\text{Types of system events}}$, the ID of its source node (i.e. head) $srcnid$, the ID of its destination node (i.e. tail) $dstnid$, a timestamp $t$, and other optional attributes $nattr^*$ depending on the data provided by each dataset. 
Edge type $etype$, source node id $srcnid$, destination node id $dstnid$, timestamp $t$ in edges correspond to operation $o$, subject $u$, object $v$, and timestamp $t$ in events, respectively. Details of attributes we consider for our model will be provided in section~\ref{sec:prvgraph_con}.

Following that, a provenance graph $PG$ can be formulated as $PG = \braket{V, E}$ where $V$ and $E$ are set of nodes and edges in the provenance graph. Figure~\ref{fig:pg_example} shows an example of a provenance graph.

\subsection{Graph Distillation}

When modeling logs, handling large datasets can be challenging, as they often contain millions of records capturing events over time. These massive datasets can create significant issues with storage, computational efficiency, and scalability during model training. This work aims to enhance scalability in processing the massive data generated by logs.

A promising deep learning approach to handle large datasets is Dataset Distillation (DD)~\cite{DDSurvey_2024}, which creates smaller, manageable synthetic datasets from the original data. 
Given a large training dataset $\mathcal{T} = \Set{(x_i, y_i)}^m_{i=1}$ of size $m$ where $x_i$ is an input data and $y_i$ is the corresponding label, dataset distillation algorithm extracts a small synthetic dataset $\mathcal{S} = \Set{(s_j, y_j)}^n_{j=1}$ of size $n$ while satisfying the two objectives specified below \cite{DDSurvey_2024}:
\begin{itemize}
    \item The size of the distilled dataset is much smaller than that of the source dataset, i.e., $n \ll m$. This ratio can be controlled by reduction rate parameter $r$.
    \item The model trained on the synthetic dataset can achieve comparable performance to the original dataset, which can be formalized as bellow \cite{GC_SGDD_2024}:
\end{itemize}
\begin{equation}
    \mathbb{E}[\ell(f_{alg(\mathcal{T})}(x), y)] \simeq \mathbb{E}[\ell(f_{alg(\mathcal{S})}(x), y)]
\end{equation}
where $f_{alg(\mathcal{T})}(x)$ denotes the output of the model $f$ trained on the dataset $\mathcal{T}$ on input $x$, and $\ell$ is the loss between predicted output and ground truth label.

Graph structure data represents relationships between events or entities found in system logs, referred to as provenance graphs. These graphs can contain thousands of nodes and millions of edges. Unlike typical dataset condensation, condensing graph structures must preserve meaningful relationships between entities and events, as shown in the original graph. Recent research highlights promising results in extending dataset distillation techniques to graph data, a technique called graph distillation~\cite{gd_GCond_2022}. Given a graph dataset $\mathcal{T} = \Set{X, A, Y}$ and synthetic graph datasets $\mathcal{S} = \Set{X', A', Y'}$, where $X$, $A$, and $Y$ are node features, adjacency matrix, and node labels, respectively, and $|X|<|X'|$. To manage meaningful representation for the synthetic graph $\mathcal{S}$ based on the original graph $\mathcal{T}$, graph condensation performs an optimization process using the GNN model $\textup{GNN}_\theta$, parameterized by $\theta$. This model learns representation on both graphs and minimizes the loss function calculated for each graph, defined as follows:

\begin{equation} 
\begin{aligned} 
\mathcal{L}^\mathcal{T} (\theta) &= \ell \left(\textup{GNN}_\theta(\mathcal{T}), Y\right), \ \mathcal{L}^\mathcal{S} (\theta) &= \ell \left(\textup{GNN}_\theta(\mathcal{S}), Y'\right), \end{aligned} 
\end{equation}
where $\ell$ represents the task-specific loss, such as cross-entropy for classification tasks, and $Y$ and $Y'$ are the labels for the original graph $\mathcal{T}$ and $\mathcal{S}$, respectively. The objective of graph condensation can be formulated as a two-way optimization problem, or bi-level problem, as follows:

\begin{equation}
\label{eq} 
\mathcal{S} = \min_\mathcal{S} \mathcal{L}^\mathcal{T}\left(\theta^\mathcal{S}\right) \quad \text{s.t.} \quad \theta^\mathcal{S} = \arg\min_\theta \mathcal{L}^\mathcal{S}(\theta), 
\end{equation}

The equation above describes the process of finding the optimal smaller synthetic graph $\mathcal{S}$ for training a model. The objective is to ensure that the performance, measured by the loss $\mathcal{L}^\mathcal{T}$, is as close as possible to the performance on the original large graph $\mathcal{T}$. The constraint encourages that the model parameters $\theta^\mathcal{S}$ are optimized by minimizing the loss $\mathcal{L}^\mathcal{S}$ on the smaller graph $\mathcal{S}$. This approach ensures that the smaller graph retains the important properties of the original graph.

\section{Problem Statement}
\label{sec:problem_stmnt}
For our ultimate goal of efficient APT detection, we consider node-level granularity. GraphDART aims to distinguish benign from malicious nodes. Since the original provenance graph is so large that it hinders the efficiency of APT detection, a scalable graph distillation approach is proposed to extract the necessary knowledge for APT detection from the original provenance graph into a condensed graph. Given a large provenance-graph dataset, $\mathcal{T}_{PG} = \Set{X_{PG}, A_{PG}, Y_{PG}}^m_{i=1}$, the aim is to construct a smaller graph, $\mathcal{S}_{PG} = \Set{X'_{PG}, A'_{PG}, Y'_{PG}}^n_{i=1}$, synthesized from the large $\mathcal{T}_{PG}$, while $\mathcal{S}_{PG}$ has comparable result for APT detection task compared to the originally large $\mathcal{T}_{PG}$.

\subsection{Threat Model}
We make similar assumptions commonly used in the literature (including, e.g., \cite{AptDtc_unicorn_2020, AptDtc_atlas_2021, AptDtc_priotracker_2018, aptDtc_omegaLog_2020, aptDtc_watson_2021}), which are summarized below:

\begin{itemize}
    \item \textbf{Integrity of the kernel/operating system:} We assume the kernel is not compromised.
    \item \textbf{Tamper-proof system logs:} We assume that system logs cannot be tampered. Therefore, the provenance graph that is built based on the source logs are correct and tamper-proof as well.
    \item \textbf{Visible attack traces:} Even though the attacker is able to imitate the normal behavior of the target system with slight deviations over long term to hide the attack progress, all actions should be captured in the system logs. In this regard, attacks that cannot be captured in the system logs, such as side-channel attack, are out of the scope of this work.
\end{itemize}

\begin{figure*}[!htbp]
    \centerline{\includegraphics[width=\linewidth]{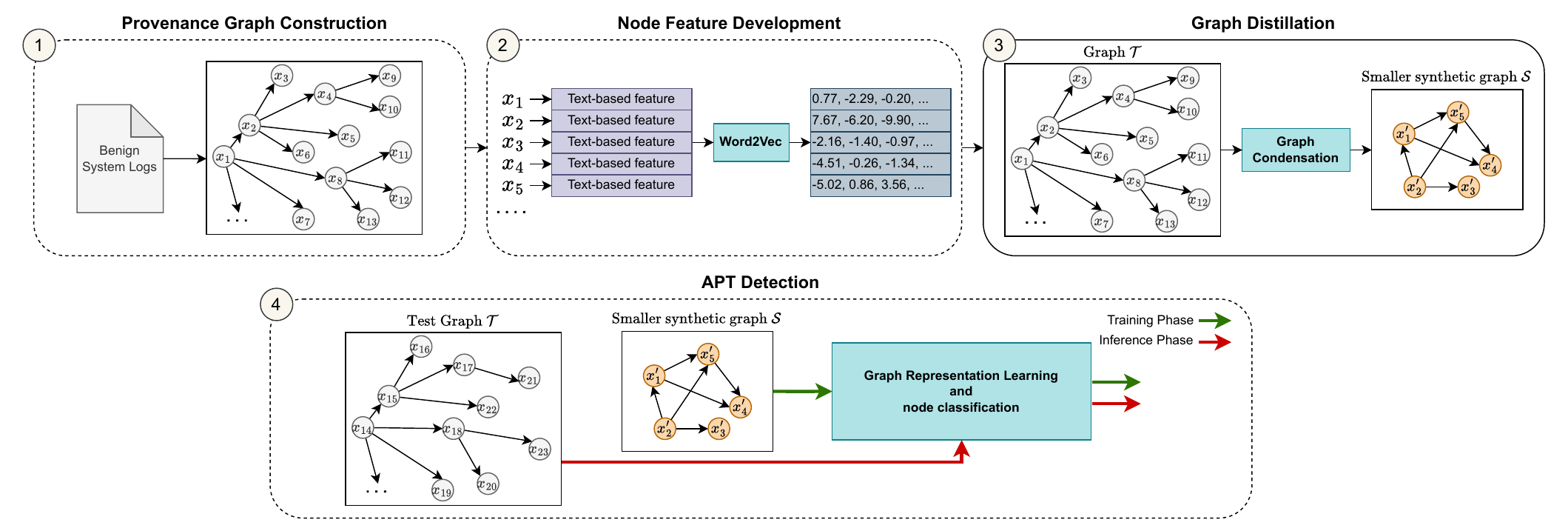}}
    \caption{Framework of GraphDART. We first create provenance graph based on the input logs (Section~\ref{sec:prvgraph_con}) and develop node features (Section~\ref{sec:node_feat_dev}). Then we apply graph distillation to get the condensed graph (Section~\ref{sec:graph_distillation}). Lastly, we train a GNN model to learn graph representation in training phase and detecting malicious nodes in inference phase (Section~\ref{sec:graph_rep_learn_node_cls})}
    \label{fig:our_framework}
\end{figure*}

\section{Design of GraphDART}
\label{sec:proposed_framework}
In this section, an overview of our GraphDART framework is first provided, and then, we explain each component of GraphDART in details.

\subsection{Overview}
Figure \ref{fig:our_framework} illustrates an overview of our framework. Given the system logs as input, GraphDART constructs the corresponding provenance graph dataset (Section \ref{sec:prvgraph_con}). Then, node features are developed based on the context information of each node and its neighbor nodes (Section~\ref{sec:node_feat_dev}). We employ strategies in FLASH~\cite{flash_2024} for constructing and developing node features of the provenance graph. Once these node features, along with the adjacency matrix and node labels, are ready, they are fed into the graph distillation module (Section~\ref{sec:graph_distillation}). This procedure is illustrated in Algorithm~\ref{alg:our_model}. The resulting condensed graph becomes the input for the graph representation learning module, which distinguishes benign nodes from malicious nodes (Section~\ref{sec:graph_rep_learn_node_cls}).

\begin{algorithm}[!htbp]
    \caption{Graph distillation procedure for APT detection}
    \label{alg:our_model}
    \renewcommand{\algorithmicrequire}{\textbf{Input:}}
    \renewcommand{\algorithmicensure}{\textbf{Output:}}
    \begin{algorithmic}[1]
        \REQUIRE System logs $LogFile$
        \ENSURE Condensed graph $\mathcal{S}_{PG}$
        
        \STATE $PG \gets ConstrucProvenanceGraph(LogFile)$

        \STATE $X_{PG} \gets DevelopNodeFeatures(PG.V)$
        \STATE $Y_{PG} \gets GetNodeTypes(PG.V)$
        \STATE $A_{PG} \gets AdjacencyMatrix(PG)$
        \STATE $\mathcal{T}_{PG} \gets (X_{PG}, A_{PG}, Y_{PG})$

        \STATE $GraphDistillMethod \gets InitGraphDistillation()$
        \STATE $\mathcal{S}_{PG} \gets GraphDistillMethod(\mathcal{T}_{PG})$

        \RETURN $\mathcal{S}_{PG}$
    \end{algorithmic}
\end{algorithm}

\subsection{Components}

\subsubsection{Provenance graph construction}
\label{sec:prvgraph_con}
Following the definitions of node and edge in Eq.~\ref{NodeDefinition} and Eq.~\ref{EdgeDefinition}, we extract nodes and edges from the input logs to construct the provenance graph similar to FLASH.
\begin{itemize}
    \item For each \textbf{node}, node ID $nid$ and node type $ntype$ are obtained from the input logs. We use process name and command-line arguments for process nodes, file path for file nodes, IP address and port for socket nodes, and module name for module nodes, as their node attributes $nattr^*$.
    \item For each \textbf{edge}, edge ID $eid$, edge type $etype$, timestamp $t$ are extracted from the input logs. Subject ID and object ID in each log record are assigned to the source node ID $srcnid$ and the destination node ID $dstnid$ of the corresponding edge, respectively.
\end{itemize}

\subsubsection{Node feature development}
\label{sec:node_feat_dev}
Like the graph construction, we define the node features procedure similar to FLASH. With the provenance graph's nodes and edges defined, we first extract text-based features for the nodes and embed them into the latent space as `node features'. We aggregate each node's attributes and the types of edges connected to it, i.e., edges linking the node to its 1-hop neighbors. These edges are sorted by their timestamps before inclusion in the node features. Incorporating neighborhood information into node features enriches the context for the ultimate task, i.e., APT detection through node classification.

After concatenating node attributes and the types of their connected edges, each resulting sentence is encoded into a dense vector using a Word2Vec model~\cite{word2vec_model_2013} trained on benign system logs. Since the standard Word2Vec model does not preserve the consecutiveness of words in a sentence, positional encoding~\cite{attention_transformers_2017} is used to capture positional attributes. Adding the Word2Vec embedding vector and the positional encoding of each node's text-based features yields a numerical representation of the node features. Finally, an averaging operation is applied to these vectors to produce a fixed-length feature vector for each node.

\subsubsection{Graph distillation}
\label{sec:graph_distillation}

Having nodes and edges, adjacency matrix $A_{PG}$ can be computed, and from the previous component, we have node features $X_{PG}$. The type of each node is considered as its label, allowing us to construct node labels $Y_{PG}$ from the node types. Together, these elements form the large graph dataset $\mathcal{T}_{PG}$. As discussed earlier, utilizing such a large graph is time-consuming and computationally inefficient. To address this issue, we apply graph distillation to obtain a condensed graph $\mathcal{S}_{PG}$. The  reduction rate parameter $r$ controls the size of the distilled graph. For instance, a reduction rate of $r=0.01$ implies that the the distilled graph contain approximately 1\% of the nodes compare to the original graph.

If a class, i.e., node type, accounts for less that 1\% of the total nodes in the original graph, we consider that class and its instances as outlier and exclude them from the graph prior to applying graph distillation. Related observations are detailed in Section~\ref{sec:obsrv_dataset}. The reason for this choice is that GraphDART aims to unify its distillation module for various distillation methods while being capable of working with even small reduction rates, $\Set{r | r \le 0.01}$. When using small values of $r$, modern methods such as GCDM~\cite{gd_gcdm_2022} and SGDD~\cite{GC_SGDD_2024} preserve the distribution of instances within each class. As a result, they tend to lose instances from classes where the number of instances is smaller than $r \nolinebreak \times \nolinebreak \textit{total number of nodes in the graph}$. In contrast, classic methods like KCenter~\cite{gd_kcenter_2018} and Herding~\cite{gd_herding_2009} do not account for the distribution of instances in the same way. By choosing a threshold of 1\% for $r$, we ensure that the framework treats all distillation models consistently.

GraphDART provides a modular framework that supports the integration of various graph distillation methods. To demonstrate its practicality, GraphDART incorporates three core-set selection methods, namely, random~\cite{gcbench_2024}, herding~\cite{gd_herding_2009}, and kcenter~\cite{gd_kcenter_2018}, one distribution matching method, namely, GCDM~\cite{gd_gcdm_2022}, and two gradient matching methods, namely, GCond~\cite{GC_SGDD_2024} and SGDD~\cite{GC_SGDD_2024}. Through extensive experiments, we studied the performance of all these distillation methods within our framework.

\subsubsection{APT detection}
\label{sec:graph_rep_learn_node_cls}

The resulting condensed graph $\mathcal{S}_{PG}$ serves as a representative of the original benign graph. For learning graph representation, we employ the GNN model of FLASH. During the training phase of the APT detector, $\mathcal{S}_{PG}$ is provided as input to the FLASH GNN model. Since the node types are considered as node labels $Y_{PG}$, the GNN learns the representation of benign nodes and predicts their types. 
The majority of benign nodes have similar neighborhood characteristics, while malicious nodes deviate from this pattern. As a result, the GNN correctly predicts benign nodes but misclassifies malicious ones.

Because the size of the condensed graph dataset $\mathcal{S}_{PG}$ is reduced by a factor of $r$ compared to the original graph dataset $\mathcal{T}_{PG}$, the GNN training is faster with the distilled graph. In the inference phase, as in FLASH, the prediction of the trained GNN for test input graph data $\mathcal{T}'$ are evaluated against the actual node labels in the ground truth, and the results are reported in Section~\ref{sec:dtc_performance_eval}. 
We compare GraphDART with FLASH specifically, as it outperforms existing APT detection techniques such as ThreaTrace~\cite{aptDtc_ThreaTrace_2022} and Unicorn~\cite{AptDtc_unicorn_2020}. For FLASH, we used the code provided by the authors and followed the default hyperparameters mentioned in their paper, but some discrepancies in the results were observed and we were unable to replicate exactly the original results presented in FLASH. Details will be provided in the next section.

\begin{table}[!htbp]
\centering
\renewcommand{\arraystretch}{1.5}
\caption{Statistics of datasets.}
\begin{tabular}{|>{\centering\arraybackslash}p{0.2\linewidth}|>{\centering\arraybackslash}p{0.23\linewidth}|>{\centering\arraybackslash}p{0.18\linewidth}|>{\centering\arraybackslash}p{0.2\linewidth}|}
\hline
Dataset & Subset & Total number of nodes & Total number of edges \\
\hline
\multirow{2}{*}{\shortstack{DARPA TC E3 \\ Cadets}} & Training phase & 362,645 & 2,059,154 \\
\cline{2-4}
 & Evaluation phase & 357,174 & 2,097,882 \\
\hline
\multirow{2}{*}{\shortstack{DARPA TC E3 \\ Theia}} & Training phase & 304,973 & 9,388,489 \\
\cline{2-4}
 & Evaluation phase & 344,768 & 9,426,694 \\
\hline
\multirow{2}{*}{\shortstack{DARPA TC E3 \\ FiveDirections}} & Training phase & 21,826 & 1,072,754 \\
\cline{2-4}
 & Evaluation phase & 21,676 & 880,826 \\
\hline
\end{tabular}
\label{tbl:darpa_graphs_info}
\end{table}

\begin{figure}[!htbp]
    \centerline{\includegraphics[width=\linewidth]{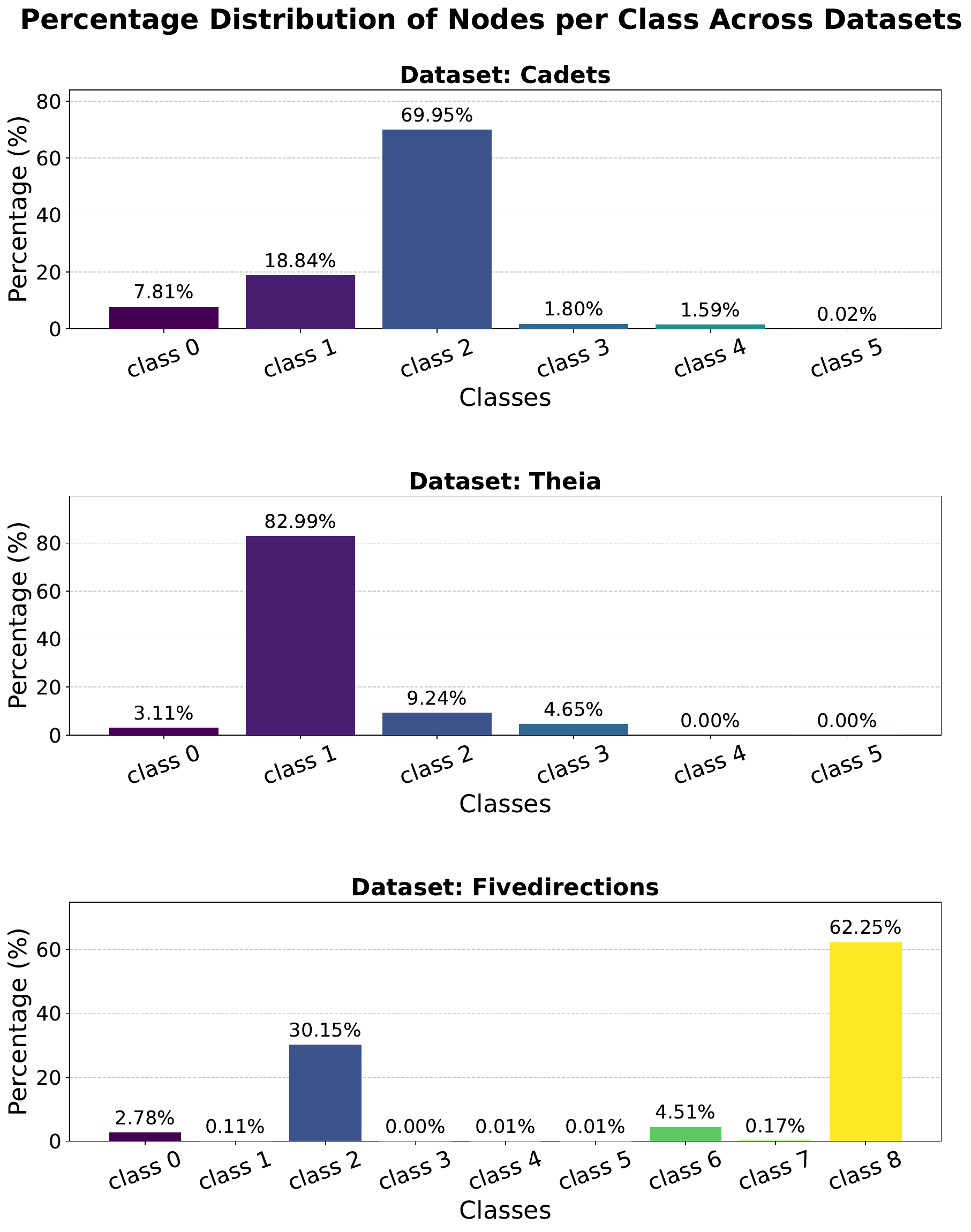}}
    \caption{Node distribution (percentage) across classes in the DARPA datasets. Table~\ref{tbl:obsrv_datasets} provides more details on the node classes.}
\label{fig:obsrv_datasets}
\end{figure}

\section{Evaluation}
\label{sec:evaluation}

We evaluated multiple versions of GraphDART and the baseline method on a machine configured with 16 vCPUs, an NVIDIA T4 GPU, 64 GB of RAM, and running Ubuntu 22.04 LTS. The implementation was carried out in Python 3, leveraging tools and libraries such as the FLASH tool, GC-Bench~\cite{gcbench_2024}, NumPy, PyTorch, Gensim, and PyTorch Geometric. This setup ensured a consistent and reliable environment for conducting experiments and comparing performance across methods. 

We introduce several variants of GraphDART in which we bring different graph distillation strategies into our framework. These methods are listed bellow:
\begin{itemize}
    \item GraphDART\_random: Our GraphDART approach in which the random~\cite{gcbench_2024} algorithm is used.
    \item GraphDART\_herding: Our GraphDART approach where we utilize the herding~\cite{gd_herding_2009} method.
    \item GraphDART\_kcenter: Our GraphDART approach that applies kcenter~\cite{gd_kcenter_2018} algorithm in the distillation module.
    \item GraphDART\_gcdm: Our GraphDART approach that produces condensed graph using GCDM~\cite{gd_gcdm_2022} method.
    \item GraphDART\_gcond: Our GraphDART approach in which the GCond~\cite{gd_GCond_2022} algorithm is used for graph distillation.
    \item GraphDART\_sgdd: Our GraphDART approach where we apply SGDD~\cite{GC_SGDD_2024} on graph for distillation.
\end{itemize}

\begin{table*}[!ht]
\centering
\renewcommand{\arraystretch}{1.5}
\caption{Description of classes and number of nodes per each class (node type) in benign graphs in DARPA datasets.}
\begin{tabular}{|>{\centering\arraybackslash}p{0.1\linewidth}|>{\centering\arraybackslash}p{0.7\linewidth}|>{\centering\arraybackslash}p{0.1\linewidth}|}
\hline
Dataset & Classes & Removed \\
\hline
DARPA TC E3 Cadets & Class 0 (SUBJECT PROCESS) with 28305 samples, class 1 (FILE OBJECT FILE) with 68331 samples, class 2 (FILE OBJECT UNIX SOCKET) with 253667 samples, class 3 (Unnamed Pipe Object) with 6518 samples, class 4 (Net Flow Object) with 5761 samples, class 5 (FILE OBJECT DIR) with 63 samples. & Class 5 \\
\hline
DARPA TC E3 Theia & Class 0 (SUBJECT PROCESS) with 9488 samples(nodes), class 1 (Memory Object) with 253105 samples, class 2 (FILE OBJECT BLOCK) with 28190 samples, class 3 (Net Flow Object) with 14189 samples, class 4 (PRINCIPAL REMOTE) with 1 sample, class 5 (PRINCIPAL LOCAL) with no sample. & Class 4 and class 5 \\
\hline
DARPA TC E3 Fivedirections & Class 0 (SUBJECT PROCESS) with 606 samples, class 1 (FILE OBJECT CHAR) with 23 samples, class 2 (VALUE TYPE SRC) with 6581 samples, class 3 (SRCSINK DATABASE) with 1 sample, class 4 (FILE OBJECT UNIX SOCKET) with 3 samples, class 5 (FILE OBJECT BLOCK) with 3 samples, class 6 (Net Flow Object) with 985 samples, class 7 (SRCSINK PROCESS MANAGEMENT) with 37 samples, class 8 (SUBJECT THREAD) with 13587 samples. & Classes 1, 3, 4, 5, and 7. \\
\hline
\end{tabular}
\label{tbl:obsrv_datasets}
\end{table*}

\begin{figure*}[!htbp]
    \centerline{\includegraphics[width=\linewidth]{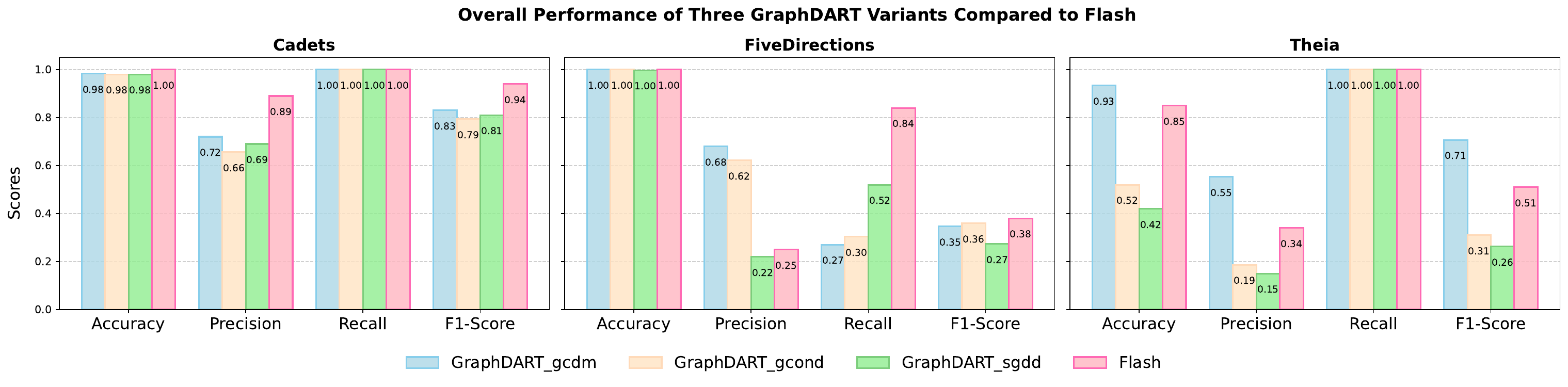}}
    \caption{APT detection performance across DARPA TC E3 datasets. GraphDART produces comparable results with FLASH while using small condensed graphs. Average results with $r \in \Set{0.006, 0.004, 0.002}$ are shown in the figure.}
    \label{fig:obsrv_prfmnc_datasets}
\end{figure*}

\subsection{Datasets and Analysis}
\label{sec:obsrv_dataset}
All methods are evaluated using three publicly-available datasets published by DARPA Transparent Computing (TC) engagement 3 (E3)~\cite{dataset_darpa3}. Table~\ref{tbl:darpa_graphs_info} shows the number of nodes and edges of graphs extracted from each dataset and enables us to evaluate the scalability of our model(s). Labels from ThreaTrace~\cite{aptDtc_ThreaTrace_2022} are utilized as ground truth.

Using DARPA datasets allows us to evaluate our proposed framework with system logs from various operating systems. Specifically, Cadets dataset includes logs from FreeBSD systems, Theia dataset is derived from a Linux-based system (Ubuntu 12.04), and the Fivedirections dataset contains logs from Windows 10 systems. These diverse system logs serve as input for our model, enabling a comprehensive assessment across different environments.

Figure~\ref{fig:obsrv_datasets} depicts the distribution of nodes across various classes in the benign graphs in the DARPA datasets, highlighting the diversity of node types present in system logs from Theia, Cadets, and Fivedirections. The actual labels for each class are introduced in Table~\ref{tbl:obsrv_datasets}. As shown, class 2 (labeled as 'FILE OBJECT UNIX SOCKET') in Cadets, class 1 (memory object labeled as 'FILE OBJECT FILE') in Theia, and classes 8 (labeled as 'SUBJECT THREAD') and 2 (labeled as 'VALUE TYPE SRC') together in Fivedirections dominate the distribution of the datasets, while the remaining classes have significantly fewer node instances.

As mentioned in Section~\ref{sec:graph_distillation}, we removed classes with less than 1\% samples before graph distillation as detailed in Table~\ref{tbl:obsrv_datasets}.

\subsection{Metrics}
We evaluated performance of APT detection with standard metrics, namely, accuracy ($\frac{TP + TN}{TP + TN + FP + FN}$), precision ($\frac{TP}{TP + FP}$), recall ($\frac{TP}{TP + FN}$), and F1-score ($2 \cdot \frac{\text{Precision} \cdot \text{Recall}}{\text{Precision} + \text{Recall}}$), using the following equations when $TP$, $FP$, $FN$, and $TN$ are number of true positives, false positives, false negatives, and true negatives, respectively.

\subsection{APT Detection Performance}
\label{sec:dtc_performance_eval}

To demonstrate the effectiveness of GraphDART across different datasets, we evaluated its detection performance using three reduction parameters: $r \in \Set{0.006, 0.004, 0.002}$. Figure~\ref{fig:obsrv_prfmnc_datasets} shows the average results of applying the three reduction parameters, compared to FLASH, which detects APTs using the original large graph. As seen, GraphDART performs competitively with our APT detection baseline, even when using the small condensed graph. 
We observed that recall is nearly  perfect (since the numbers are represented as rounded two-digit decimal numbers) for most cases due to the low number of false negatives. It is important noting that the numbers are rounded by two-digit decimals, so high number of true positives with low number of false negatives leads to a nearly perfect recall. However, a moderate number of false positives decreased the precision in all methods.

Tables~\ref{tbl:performance_cadets}, \ref{tbl:performance_theia}, and \ref{tbl:performance_fived} show the results of the study. The numbers are rounded to be represented as two-digit floating point numbers. We can interpret the results from two perspectives: first, studying the choice of reduction parameter $r$ and its effect on detection performance with diverse distillation methods, and second, evaluating the choice of distillation method and its effect of detection performance.

\begin{table}[!htbp]
\renewcommand{\arraystretch}{1.1}
\centering
\caption{Detection performance (Accuracy (Ac.), Precision (Pr.), Recall (Re.), and F1-Score (F1)) on DARPA E3 Cadets datasets using GraphDART: a.~GraphDART\_random, b.~GraphDART\_herding, c.~GraphDART\_kcenter, d.~GraphDART\_gcdm, e.~GraphDART\_gcond, and f.~GraphDART\_sgdd. Training time of the graph representation leaning model (t) is reported in seconds. $r$ is the reduction parameter.}
\label{tbl:performance_cadets}
\begin{tabular}{@{\extracolsep{1pt}}c|c|cccccc|c}
\cline{2-8}
 & \multicolumn{7}{c|}{Various GraphDART versions} & \\ \cline{2-9}
 & $r$ & a & b & c & d & e & f & \multicolumn{1}{c|}{\makecell{FLASH}} \\ \hline

\multicolumn{1}{|c|}{\multirow{7}{*}{Ac.}} & 0.05           & 0.29   & 0.97    & 0.94    & -    & -     & -    & \multicolumn{1}{c|}{\multirow{7}{*}{1}}    \\
\multicolumn{1}{|c|}{}                          & 0.03           & 0.29   & 0.29    & 0.21    & -    & -     & -    & \multicolumn{1}{c|}{}                      \\
\multicolumn{1}{|c|}{}                          & 0.01           & 0.97   & 0.27    & 0.22    & -    & -     & -    & \multicolumn{1}{c|}{}                      \\
\multicolumn{1}{|c|}{}                          & 0.008          & 0.28   & 0.99    & 0.98    & 0.97 & 1     & -    & \multicolumn{1}{c|}{}                      \\
\multicolumn{1}{|c|}{}                          & 0.006          & 0.94   & 0.95    & 0.34    & 0.97 & 0.98  & 0.96 & \multicolumn{1}{c|}{}                      \\
\multicolumn{1}{|c|}{}                          & 0.004          & 0.98   & 0.3     & 0.97    & 0.99 & 0.98  & 0.98 & \multicolumn{1}{c|}{}                      \\
\multicolumn{1}{|c|}{}                          & 0.002          & 0.27   & 0.2     & 0.94    & 0.99 & 0.98  & 1    & \multicolumn{1}{c|}{}                      \\ \hline
\multicolumn{1}{|c|}{\multirow{7}{*}{Pr.}} & 0.05           & 0.05   & 0.54    & 0.45    & -    & -     & -    & \multicolumn{1}{c|}{\multirow{7}{*}{0.89}} \\
\multicolumn{1}{|c|}{}                          & 0.03           & 0.05   & 0.05    & 0.05    & -    & -     & -    & \multicolumn{1}{c|}{}                      \\
\multicolumn{1}{|c|}{}                          & 0.01           & 0.61   & 0.05    & 0.05    & -    & -     & -    & \multicolumn{1}{c|}{}                      \\
\multicolumn{1}{|c|}{}                          & 0.008          & 0.05   & 0.79    & 0.68    & 0.57 & 0.9   & -    & \multicolumn{1}{c|}{}                      \\
\multicolumn{1}{|c|}{}                          & 0.006          & 0.43   & 0.46    & 0.06    & 0.58 & 0.64  & 0.55 & \multicolumn{1}{c|}{}                      \\
\multicolumn{1}{|c|}{}                          & 0.004          & 0.63   & 0.06    & 0.6     & 0.79 & 0.64  & 0.63 & \multicolumn{1}{c|}{}                      \\
\multicolumn{1}{|c|}{}                          & 0.002          & 0.05   & 0.05    & 0.42    & 0.79 & 0.69  & 0.89 & \multicolumn{1}{c|}{}                      \\ \hline
\multicolumn{1}{|c|}{\multirow{7}{*}{Re.}}   & 0.05           & 1      & 1       & 1       & -    & -     & -    & \multicolumn{1}{c|}{\multirow{7}{*}{1}}    \\
\multicolumn{1}{|c|}{}                          & 0.03           & 1      & 1       & 1       & -    & -     & -    & \multicolumn{1}{c|}{}                      \\
\multicolumn{1}{|c|}{}                          & 0.01           & 1      & 1       & 1       & -    & -     & -    & \multicolumn{1}{c|}{}                      \\
\multicolumn{1}{|c|}{}                          & 0.008          & 1      & 1       & 1       & 1    & 1     & -    & \multicolumn{1}{c|}{}                      \\
\multicolumn{1}{|c|}{}                          & 0.006          & 1      & 1       & 1       & 1    & 1     & 1    & \multicolumn{1}{c|}{}                      \\
\multicolumn{1}{|c|}{}                          & 0.004          & 1      & 1       & 1       & 1    & 1     & 1    & \multicolumn{1}{c|}{}                      \\
\multicolumn{1}{|c|}{}                          & 0.002          & 1      & 1       & 1       & 1    & 1     & 1    & \multicolumn{1}{c|}{}                      \\ \hline
\multicolumn{1}{|c|}{\multirow{7}{*}{F1}}   & 0.05           & 0.1    & 0.71    & 0.62    & -    & -     & -    & \multicolumn{1}{c|}{\multirow{7}{*}{0.94}} \\
\multicolumn{1}{|c|}{}                          & 0.03           & 0.1    & 0.1     & 0.1     & -    & -     & -    & \multicolumn{1}{c|}{}                      \\
\multicolumn{1}{|c|}{}                          & 0.01           & 0.75   & 0.1     & 0.1     & -    & -     & -    & \multicolumn{1}{c|}{}                      \\
\multicolumn{1}{|c|}{}                          & 0.008          & 0.1    & 0.89    & 0.81    & 0.73 & 0.95  & -    & \multicolumn{1}{c|}{}                      \\
\multicolumn{1}{|c|}{}                          & 0.006          & 0.6    & 0.63    & 0.1     & 0.73 & 0.78  & 0.71 & \multicolumn{1}{c|}{}                      \\
\multicolumn{1}{|c|}{}                          & 0.004          & 0.78   & 0.1     & 0.75    & 0.88 & 0.78  & 0.78 & \multicolumn{1}{c|}{}                      \\
\multicolumn{1}{|c|}{}                          & 0.002          & 0.1    & 0.1     & 0.59    & 0.88 & 0.82  & 0.94 & \multicolumn{1}{c|}{}                      \\ \hline

\multicolumn{1}{|c|}{\multirow{7}{*}{t}} & 0.05  & 6 & 6 & 8 & -  & -  & - & \multicolumn{1}{c|}{\multirow{7}{*}{90}} \\
\multicolumn{1}{|c|}{} & 0.03  & 6 & 6 & 5 & -  & -  & - & \multicolumn{1}{c|}{} \\
\multicolumn{1}{|c|}{} & 0.01  & 5 & 5 & 6 & -  & -  & - & \multicolumn{1}{c|}{} \\
\multicolumn{1}{|c|}{} & 0.008 & 6 & 5 & 5 & 11 & 14 & - & \multicolumn{1}{c|}{} \\
\multicolumn{1}{|c|}{} & 0.006 & 6 & 6 & 6 & 7  & 9  & 7 & \multicolumn{1}{c|}{} \\
\multicolumn{1}{|c|}{} & 0.004 & 6 & 6 & 6 & 7  & 6  & 7 & \multicolumn{1}{c|}{} \\
\multicolumn{1}{|c|}{} & 0.002 & 6 & 6 & 5 & 6  & 8  & 6 & \multicolumn{1}{c|}{} \\ \hline
\end{tabular}
\end{table}

For the Cadets dataset, setting reduction rate from $\Set{r | r \in \Set{0.05, 0.03, 0.01, 0.008, 0.006, 0.004, 0.002}}$ resulted in table~\ref{tbl:performance_cadets}. As we can see, when using gradient-matching-based methods, i.e., GraphDART\_sgdd and GraphDART\_gcond, and distribution-matching-based method, i.e., GraphDART\_gcdm, GraphDART can detect malicious nodes with high accuracy similar to FLASH. However, coreset-selection-based methods, i.e., GraphDART\_random, GraphDART\_herding, and GraphDART\_kcenter, produce fluctuating results due to their nondeterministic nature.

\begin{table}[!htbp]
\renewcommand{\arraystretch}{1.1}
\centering
\caption{Detection performance (Accuracy (Ac.), Precision (Pr.), Recall (Re.), and F1-Score (F1)) on DARPA E3 Theia datasets using GraphDART: a.~GraphDART\_random, b.~GraphDART\_herding, c.~GraphDART\_kcenter, d.~GraphDART\_gcdm, e.~GraphDART\_gcond, and f.~GraphDART\_sgdd. Training time of the graph representation leaning model (t) is reported in seconds. $r$ is the reduction parameter.}

\label{tbl:performance_theia}
\begin{tabular}{@{\extracolsep{1pt}}c|c|cccccc|c}
\cline{2-8}
 & \multicolumn{7}{c|}{Various GraphDART versions} & \\ \cline{2-9}
 & $r$ & a & b & c & d & e & f & \multicolumn{1}{c|}{\makecell{FLASH}} \\ \hline

\multicolumn{1}{|c|}{\multirow{7}{*}{Ac.}} & 0.05           & 0.97   & 0.62    & 0.25    & -    & -     & -    & \multicolumn{1}{c|}{\multirow{7}{*}{0.85}} \\
\multicolumn{1}{|c|}{}                          & 0.03           & 0.83   & 0.94    & 0.35    & -    & -     & -    & \multicolumn{1}{c|}{}                      \\
\multicolumn{1}{|c|}{}                          & 0.01           & 0.92   & 0.27    & 0.55    & 0.82 & 0.39  & -    & \multicolumn{1}{c|}{}                      \\
\multicolumn{1}{|c|}{}                          & 0.008          & 0.44   & 0.83    & 0.37    & 0.2  & 0.4   & -    & \multicolumn{1}{c|}{}                      \\
\multicolumn{1}{|c|}{}                          & 0.006          & 0.87   & 0.82    & 0.26    & 0.96 & 0.68  & 0.73 & \multicolumn{1}{c|}{}                      \\
\multicolumn{1}{|c|}{}                          & 0.004          & 0.43   & 0.96    & 0.18    & 0.95 & 0.75  & 0.32 & \multicolumn{1}{c|}{}                      \\
\multicolumn{1}{|c|}{}                          & 0.002          & 0.53   & 0.36    & 0.12    & 0.89 & 0.13  & 0.21 & \multicolumn{1}{c|}{}                      \\ \hline
\multicolumn{1}{|c|}{\multirow{7}{*}{Pr.}} & 0.05           & 0.69   & 0.18    & 0.12    & -    & -     & -    & \multicolumn{1}{c|}{\multirow{7}{*}{0.34}} \\
\multicolumn{1}{|c|}{}                          & 0.03           & 0.32   & 0.57    & 0.13    & -    & -     & -    & \multicolumn{1}{c|}{}                      \\
\multicolumn{1}{|c|}{}                          & 0.01           & 0.51   & 0.12    & 0.18    & 0.3  & 0.14  & -    & \multicolumn{1}{c|}{}                      \\
\multicolumn{1}{|c|}{}                          & 0.008          & 0.13   & 0.33    & 0.14    & 0.11 & 0.14  & -    & \multicolumn{1}{c|}{}                      \\
\multicolumn{1}{|c|}{}                          & 0.006          & 0.37   & 0.3     & 0.12    & 0.64 & 0.21  & 0.22 & \multicolumn{1}{c|}{}                      \\
\multicolumn{1}{|c|}{}                          & 0.004          & 0.15   & 0.64    & 0.11    & 0.6  & 0.24  & 0.12 & \multicolumn{1}{c|}{}                      \\
\multicolumn{1}{|c|}{}                          & 0.002          & 0.18   & 0.13    & 0.11    & 0.42 & 0.11  & 0.11 & \multicolumn{1}{c|}{}                      \\ \hline
\multicolumn{1}{|c|}{\multirow{7}{*}{Re.}}   & 0.05           & 1      & 1       & 1       & -    & -     & -    & \multicolumn{1}{c|}{\multirow{7}{*}{1}}    \\
\multicolumn{1}{|c|}{}                          & 0.03           & 1      & 1       & 1       & -    & -     & -    & \multicolumn{1}{c|}{}                      \\
\multicolumn{1}{|c|}{}                          & 0.01           & 1      & 1       & 1       & 1    & 1     & -    & \multicolumn{1}{c|}{}                      \\
\multicolumn{1}{|c|}{}                          & 0.008          & 1      & 1       & 1       & 1    & 1     & -    & \multicolumn{1}{c|}{}                      \\
\multicolumn{1}{|c|}{}                          & 0.006          & 1      & 1       & 1       & 1    & 1     & 1    & \multicolumn{1}{c|}{}                      \\
\multicolumn{1}{|c|}{}                          & 0.004          & 1      & 1       & 1       & 1    & 1     & 1    & \multicolumn{1}{c|}{}                      \\
\multicolumn{1}{|c|}{}                          & 0.002          & 1      & 1       & 1       & 1    & 1     & 1    & \multicolumn{1}{c|}{}                      \\ \hline
\multicolumn{1}{|c|}{\multirow{7}{*}{F1}}   & 0.05           & 0.82   & 0.31    & 0.21    & -    & -     & -    & \multicolumn{1}{c|}{\multirow{7}{*}{0.51}} \\
\multicolumn{1}{|c|}{}                          & 0.03           & 0.48   & 0.73    & 0.24    & -    & -     & -    & \multicolumn{1}{c|}{}                      \\
\multicolumn{1}{|c|}{}                          & 0.01           & 0.68   & 0.22    & 0.3     & 0.46 & 0.24  & -    & \multicolumn{1}{c|}{}                      \\
\multicolumn{1}{|c|}{}                          & 0.008          & 0.24   & 0.49    & 0.24    & 0.2  & 0.25  & -    & \multicolumn{1}{c|}{}                      \\
\multicolumn{1}{|c|}{}                          & 0.006          & 0.54   & 0.47    & 0.21    & 0.78 & 0.35  & 0.37 & \multicolumn{1}{c|}{}                      \\
\multicolumn{1}{|c|}{}                          & 0.004          & 0.26   & 0.78    & 0.2     & 0.75 & 0.39  & 0.22 & \multicolumn{1}{c|}{}                      \\
\multicolumn{1}{|c|}{}                          & 0.002          & 0.3    & 0.24    & 0.19    & 0.59 & 0.19  & 0.2  & \multicolumn{1}{c|}{}                      \\ \hline

\multicolumn{1}{|c|}{\multirow{7}{*}{t}} & 0.05  & 6 & 6 & 6 & -  & - & - & \multicolumn{1}{c|}{\multirow{7}{*}{282}} \\
\multicolumn{1}{|c|}{} & 0.03  & 6 & 6 & 6 & -  & - & - & \multicolumn{1}{c|}{} \\
\multicolumn{1}{|c|}{} & 0.01  & 6 & 6 & 6 & 12 & 9 & - & \multicolumn{1}{c|}{} \\
\multicolumn{1}{|c|}{} & 0.008 & 6 & 6 & 5 & 7  & 8 & - & \multicolumn{1}{c|}{} \\
\multicolumn{1}{|c|}{} & 0.006 & 6 & 6 & 6 & 7  & 9 & 7 & \multicolumn{1}{c|}{} \\
\multicolumn{1}{|c|}{} & 0.004 & 6 & 6 & 6 & 7  & 6 & 6 & \multicolumn{1}{c|}{} \\
\multicolumn{1}{|c|}{} & 0.002 & 6 & 5 & 5 & 7  & 6 & 5 & \multicolumn{1}{c|}{} \\ \hline
\end{tabular}
\end{table}

In Table~\ref{tbl:performance_theia}, which is related to Theia dataset, GraphDART\_sgdd, GraphDART\_gcond, and GraphDART\_gcdm exhibited a peak performance at $r=0.006$, $r=0.004$, and $r=0.006$, respectively. This highlights the importance of choosing appropriate reduction rate for graph distillation based on the application. Similar to the observation on Cadets dataset, detection performance fluctuates when employing coreset methods in the distillation module. For both Table~\ref{tbl:performance_cadets} and Table~\ref{tbl:performance_theia}, GraphDART and FLASH reach nearly perfect recall because of the low number of false negatives.

\begin{table}[!htbp]
\renewcommand{\arraystretch}{1.1}
\centering
\caption{Detection performance (Accuracy (Ac.), Precision (Pr.), Recall (Re.), and F1-Score (F1)) on DARPA E3 Fivedirections datasets using GraphDART: a.~GraphDART\_random, b.~GraphDART\_herding, c.~GraphDART\_kcenter, d.~GraphDART\_gcdm, e.~GraphDART\_gcond, and f.~GraphDART\_sgdd. Training time of the graph representation leaning model (t) is reported in seconds. $r$ is the reduction parameter.}
\label{tbl:performance_fived}
\begin{tabular}{@{\extracolsep{1pt}}c|c|cccccc|c}
\cline{2-8}
 & \multicolumn{7}{c|}{Various GraphDART versions} & \\ \cline{2-9}
 & $r$ & a & b & c & d & e & f & \multicolumn{1}{c|}{\makecell{FLASH}} \\ \hline

\multicolumn{1}{|c|}{\multirow{7}{*}{Ac.}} & 0.05           & 1      & 0.99    & 1       & 1    & 1     & 1    & \multicolumn{1}{c|}{\multirow{7}{*}{1}}    \\
\multicolumn{1}{|c|}{}                          & 0.03           & 0.99   & 1       & 1       & 1    & 1     & 1    & \multicolumn{1}{c|}{}                      \\
\multicolumn{1}{|c|}{}                          & 0.01           & 0.99   & 1       & 1       & 0.99 & 1     & 1    & \multicolumn{1}{c|}{}                      \\
\multicolumn{1}{|c|}{}                          & 0.008          & 1      & 0.99    & 1       & 1    & 1     & 1    & \multicolumn{1}{c|}{}                      \\
\multicolumn{1}{|c|}{}                          & 0.006          & 0.99   & 1       & 1       & 1    & 1     & 1    & \multicolumn{1}{c|}{}                      \\
\multicolumn{1}{|c|}{}                          & 0.004          & 1      & 1       & 1       & 1    & 1     & 0.99 & \multicolumn{1}{c|}{}                      \\
\multicolumn{1}{|c|}{}                          & 0.002          & 0.99   & 0.99    & 1       & 1    & 1     & 1    & \multicolumn{1}{c|}{}                      \\ \hline
\multicolumn{1}{|c|}{\multirow{7}{*}{Pr.}} & 0.05           & 0.22   & 0.17    & 0.91    & 0.91 & 0.96  & 0.19 & \multicolumn{1}{c|}{\multirow{7}{*}{0.25}} \\
\multicolumn{1}{|c|}{}                          & 0.03           & 0.17   & 0.57    & 0.74    & 0.2  & 0.9   & 0.94 & \multicolumn{1}{c|}{}                      \\
\multicolumn{1}{|c|}{}                          & 0.01           & 0.15   & 0.94    & 0.18    & 0.17 & 0.89  & 0.58 & \multicolumn{1}{c|}{}                      \\
\multicolumn{1}{|c|}{}                          & 0.008          & 0.19   & 0.17    & 0.23    & 0.95 & 0.89  & 0.91 & \multicolumn{1}{c|}{}                      \\
\multicolumn{1}{|c|}{}                          & 0.006          & 0.17   & 0.94    & 0.93    & 0.83 & 0.95  & 0.26 & \multicolumn{1}{c|}{}                      \\
\multicolumn{1}{|c|}{}                          & 0.004          & 0.21   & 0.29    & 0.95    & 0.9  & 0.035 & 0.17 & \multicolumn{1}{c|}{}                      \\
\multicolumn{1}{|c|}{}                          & 0.002          & 0.17   & 0.17    & 0.24    & 0.31 & 0.88  & 0.23 & \multicolumn{1}{c|}{}                      \\ \hline
\multicolumn{1}{|c|}{\multirow{7}{*}{Re.}}   & 0.05           & 0.79   & 0.95    & 0.23    & 0.22 & 0.16  & 0.92 & \multicolumn{1}{c|}{\multirow{7}{*}{0.84}} \\
\multicolumn{1}{|c|}{}                          & 0.03           & 0.96   & 0.33    & 0.21    & 0.95 & 0.15  & 0.19 & \multicolumn{1}{c|}{}                      \\
\multicolumn{1}{|c|}{}                          & 0.01           & 0.78   & 0.18    & 0.31    & 0.97 & 0.19  & 0.37 & \multicolumn{1}{c|}{}                      \\
\multicolumn{1}{|c|}{}                          & 0.008          & 0.68   & 0.97    & 0.42    & 0.21 & 0.21  & 0.22 & \multicolumn{1}{c|}{}                      \\
\multicolumn{1}{|c|}{}                          & 0.006          & 0.97   & 0.15    & 0.18    & 0.23 & 0.22  & 0.28 & \multicolumn{1}{c|}{}                      \\
\multicolumn{1}{|c|}{}                          & 0.004          & 0.62   & 0.37    & 0.18    & 0.22 & 0.5   & 0.97 & \multicolumn{1}{c|}{}                      \\
\multicolumn{1}{|c|}{}                          & 0.002          & 0.94   & 0.97    & 0.79    & 0.36 & 0.19  & 0.31 & \multicolumn{1}{c|}{}                      \\ \hline
\multicolumn{1}{|c|}{\multirow{7}{*}{F1}}   & 0.05           & 0.35   & 0.29    & 0.36    & 0.36 & 0.27  & 0.32 & \multicolumn{1}{c|}{\multirow{7}{*}{0.38}} \\
\multicolumn{1}{|c|}{}                          & 0.03           & 0.29   & 0.42    & 0.33    & 0.34 & 0.26  & 0.32 & \multicolumn{1}{c|}{}                      \\
\multicolumn{1}{|c|}{}                          & 0.01           & 0.26   & 0.3     & 0.22    & 0.29 & 0.31  & 0.45 & \multicolumn{1}{c|}{}                      \\
\multicolumn{1}{|c|}{}                          & 0.008          & 0.3    & 0.29    & 0.3     & 0.35 & 0.34  & 0.35 & \multicolumn{1}{c|}{}                      \\
\multicolumn{1}{|c|}{}                          & 0.006          & 0.29   & 0.25    & 0.3     & 0.35 & 0.36  & 0.27 & \multicolumn{1}{c|}{}                      \\
\multicolumn{1}{|c|}{}                          & 0.004          & 0.31   & 0.32    & 0.31    & 0.36 & 0.41  & 0.29 & \multicolumn{1}{c|}{}                      \\
\multicolumn{1}{|c|}{}                          & 0.002          & 0.29   & 0.29    & 0.37    & 0.33 & 0.31  & 0.26 & \multicolumn{1}{c|}{}                      \\ \hline

\multicolumn{1}{|c|}{\multirow{7}{*}{t}} & 0.05  & 5 & 5 & 5 & 6 & 6 & 6 & \multicolumn{1}{c|}{\multirow{7}{*}{18}} \\
\multicolumn{1}{|c|}{} & 0.03  & 6 & 5 & 5 & 6 & 5 & 5 & \multicolumn{1}{c|}{} \\
\multicolumn{1}{|c|}{} & 0.01  & 5 & 5 & 5 & 5 & 5 & 5 & \multicolumn{1}{c|}{} \\
\multicolumn{1}{|c|}{} & 0.008 & 5 & 5 & 5 & 5 & 5 & 5 & \multicolumn{1}{c|}{} \\
\multicolumn{1}{|c|}{} & 0.006 & 5 & 5 & 5 & 6 & 5 & 5 & \multicolumn{1}{c|}{} \\
\multicolumn{1}{|c|}{} & 0.004 & 5 & 5 & 5 & 5 & 5 & 5 & \multicolumn{1}{c|}{} \\
\multicolumn{1}{|c|}{} & 0.002 & 5 & 5 & 5 & 5 & 5 & 6 & \multicolumn{1}{c|}{} \\ \hline
\end{tabular}

\end{table}

The results of our experiments on Fivedirections dataset are given in Table~\ref{tbl:performance_fived}. The nondeterministic behavior of coreset approaches persist in this results as seen with the other two datasets. Due to the high number of false positive predictions in Fivedirections dataset, precision has the most drop in this dataset compared to the others. However, GraphDART produces comparable results to the FLASH baseline.

\subsection{Runtime Performance of Graph Learning}

The graph produced by the graph distillation module serves as the input for the APT detection module during the training phase. Training with large graphs in the graph learning process requires significantly higher computational resources and time. Tables~\ref{tbl:performance_cadets}, \ref{tbl:performance_theia}, and \ref{tbl:performance_fived} illustrate the efficiency of graph learning for the APT detection task in GraphDART. To ensure a fair comparison, we employed the same number of iterations, i.e., training epochs, for graph representation learning as specified in FLASH.

In the case of smaller provenance graphs, such as the graph in the FiveDirections dataset, the training time difference between FLASH and GraphDART is minimal, approximately six seconds. However, the advantage of GraphDART becomes projected with larger graphs, such as the graph from the Theia dataset. For this dataset, GraphDART achieves a significant reduction in training time, with a difference of over four minutes. This trend is observed even with fewer than 25 iterations for graph learning across all cases. Notably, increasing the number of training epochs further amplifies this time-saving advantage.

\section{Conclusion}
\label{sec:conclusion}

In this paper, we have presented GraphDART, a modular framework for efficient and scalable APT detection using graph distillation on provenance graphs. By leveraging graph distillation techniques, we reduced the computational overhead of executing graph learning on large provenance graphs while preserving critical structural information. Through extensive experiments, we demonstrated the effectiveness of various distillation methods integrated into GraphDART and showcased its compatibility with state-of-the-art APT detection methods. Our results highlight the efficiency in training phase of APT detection and the effectiveness of GraphDART balancing efficiency and accuracy, making it a powerful solution for real-time intrusion detection in large-scale CPSSs and enterprise environments.

\section*{Acknowledgements}
The authors would like to thank Professor Zahir Tari (RMIT University) and Dr. Nasrin Sohrabi (Deakin University) for their supervision and valuable instructions on this work (in the early stage), as well as Bohan Cheng (RMIT University) for his help with the experiments.

\bibliographystyle{ieeetr}
\bibliography{references}


\begin{IEEEbiography}
[{\includegraphics[width=1in, height=1.25in, clip, keepaspectratio]{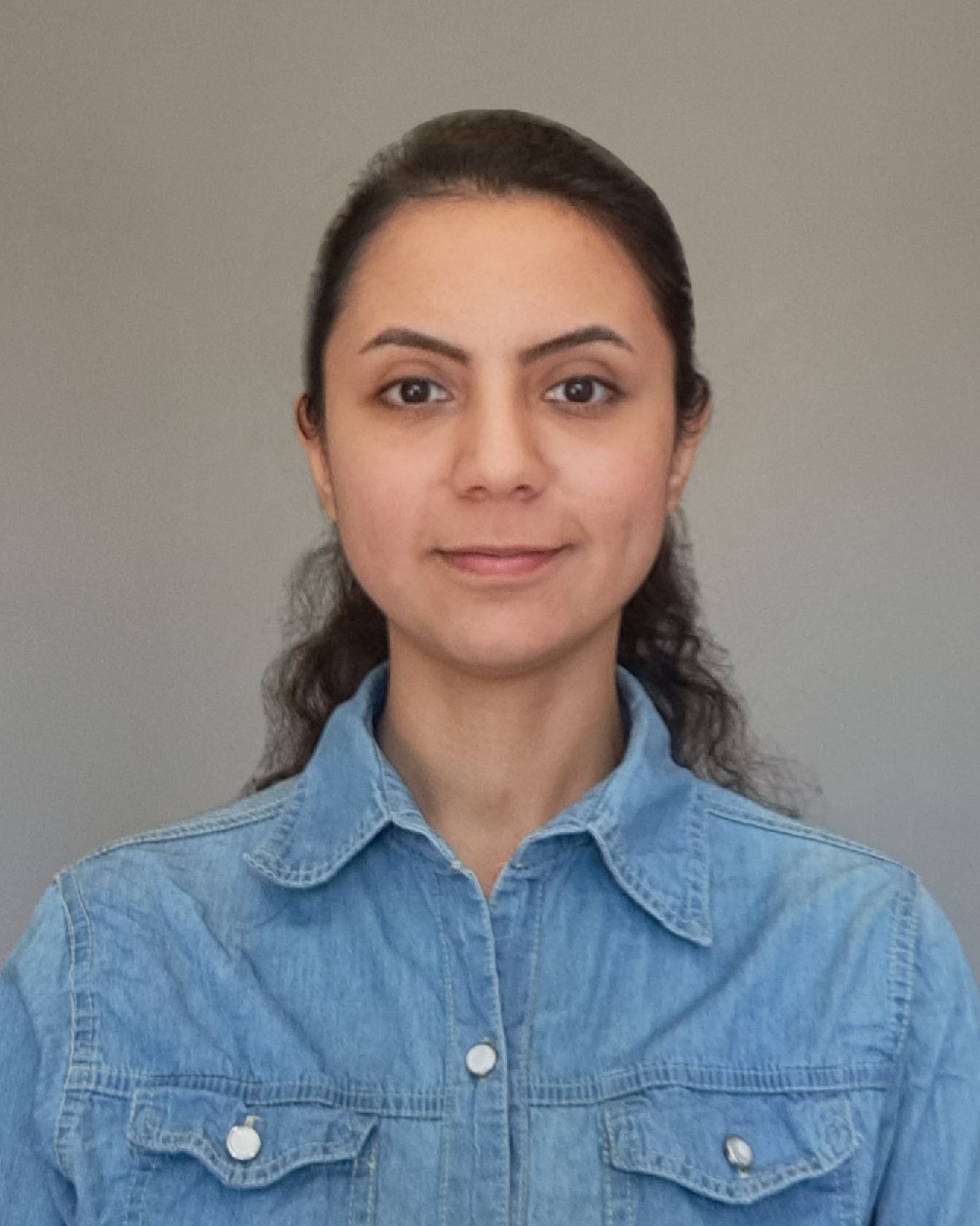}}]
{Saba Fathi Rabooki}
is a Ph.D. candidate at the School of Computing Technologies, RMIT University, Melbourne, Australia. She received her Bachelor of Science degree in Computer Engineering from K. N. Toosi University of Technology, Tehran, Iran, in 2022. Her research interests include anomaly detection, graph learning, and distributed systems.
\end{IEEEbiography}

\begin{IEEEbiography}
[{\includegraphics[width=1in, height=1.25in, clip, keepaspectratio]{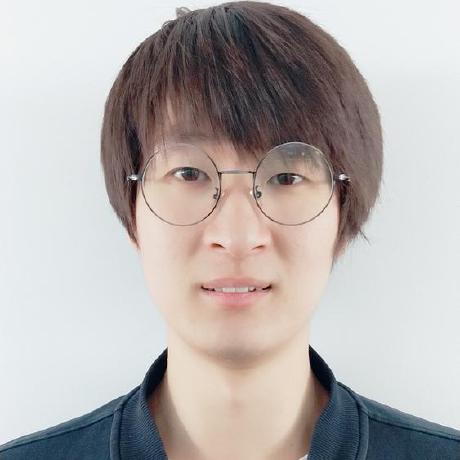}}]
{Bowen Li} received his Bachelor of computer science degree from Northeastern University, China, in 2019 and his Master of information technology degree from RMIT University, Australia, in 2024. His research interests include graph learning, artificial intelligence, cloud computing, and open-source technologies.
\end{IEEEbiography}
\vspace{30pt}

\begin{IEEEbiography}
[{\includegraphics[width=1in, height=1.25in, clip, keepaspectratio]{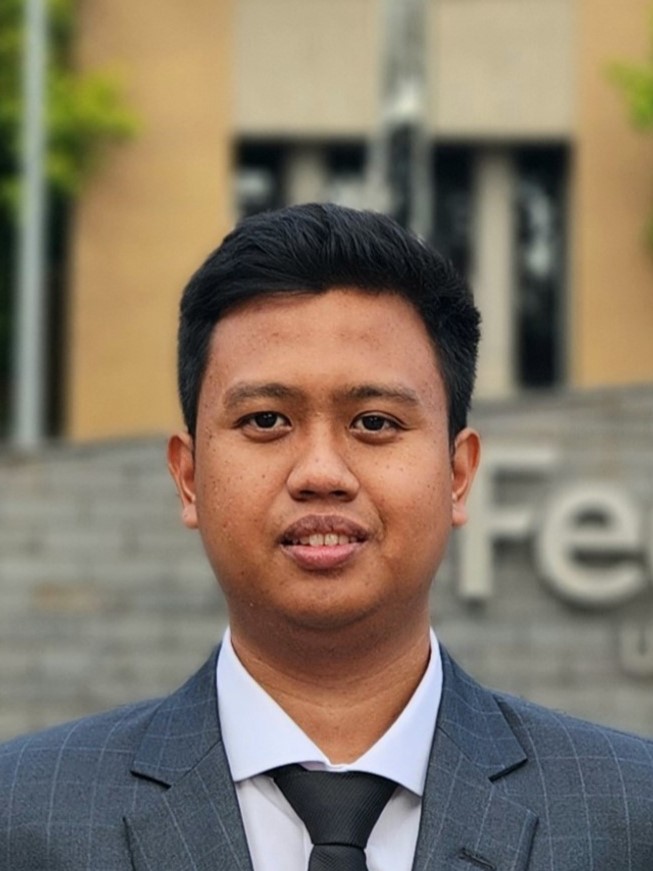}}]
{Falih Gozi Febrinanto} received his bachelor of computer science degree from the University of Brawijaya, Malang, Indonesia, in 2018 and his master of technology degree from Federation University Australia, Ballarat, Australia, in 2021. He is currently pursuing a Ph.D. degree in information technology at Federation University Australia, Ballarat, Australia. His research interests include graph learning, artificial intelligence, and anomaly detection.

\end{IEEEbiography}

\begin{IEEEbiography}
[{\includegraphics[width=1in, height=1.25in, clip, keepaspectratio]{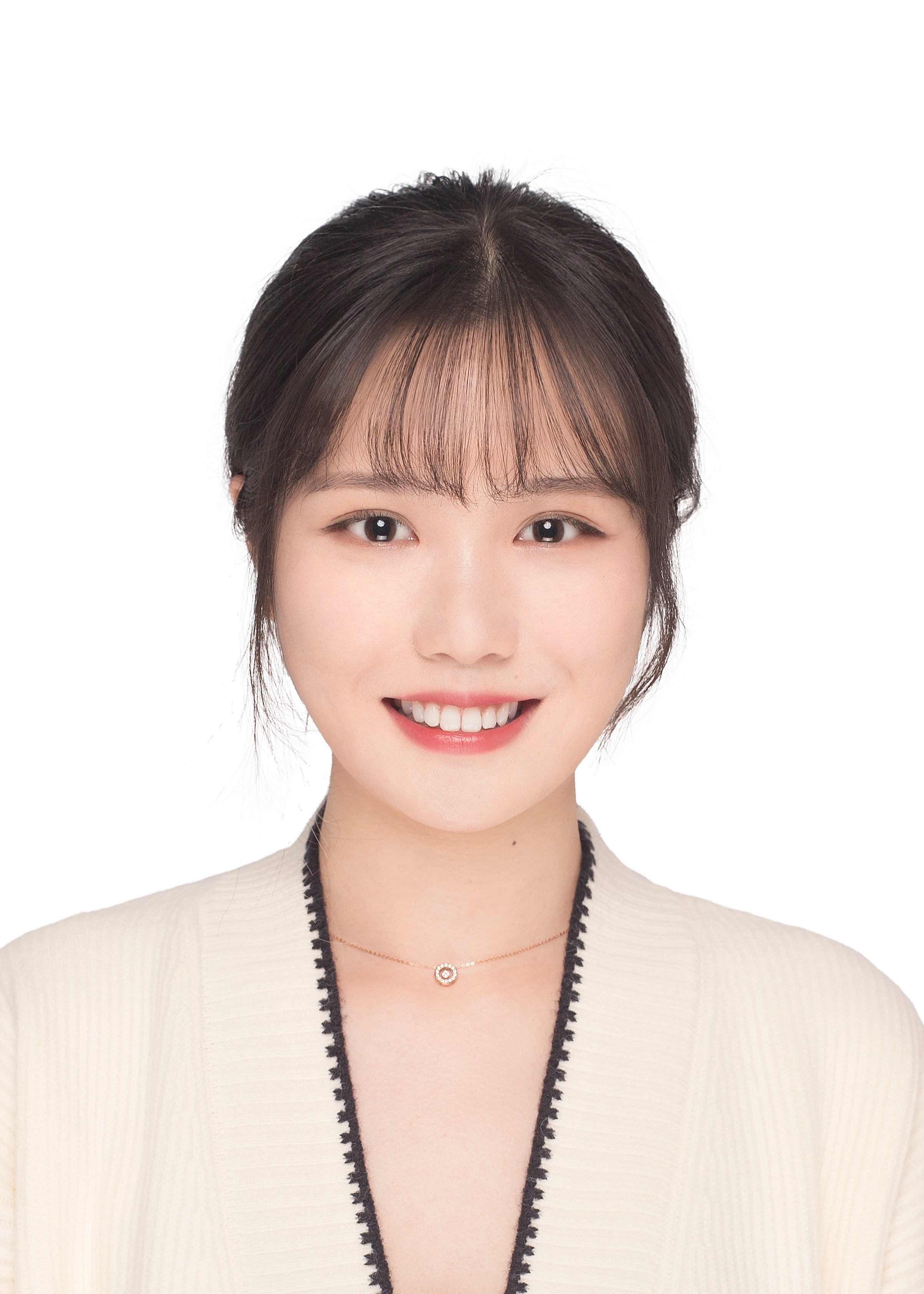}}]
{Ciyuan Peng} (Student Member, IEEE) is a Ph.D. candidate at the Institute of Innovation, Science and Sustainability, Federation University Australia. She received the B.Sc. degree from Chongqing Normal University, China, in 2018, and the M.Sc. degree from Chung-Ang University, Korea, in 2020. Her research interests include graph learning, brain science, and digital health.
\end{IEEEbiography}

\begin{IEEEbiography}
[{\includegraphics[width=1in, height=1.25in, clip, keepaspectratio]{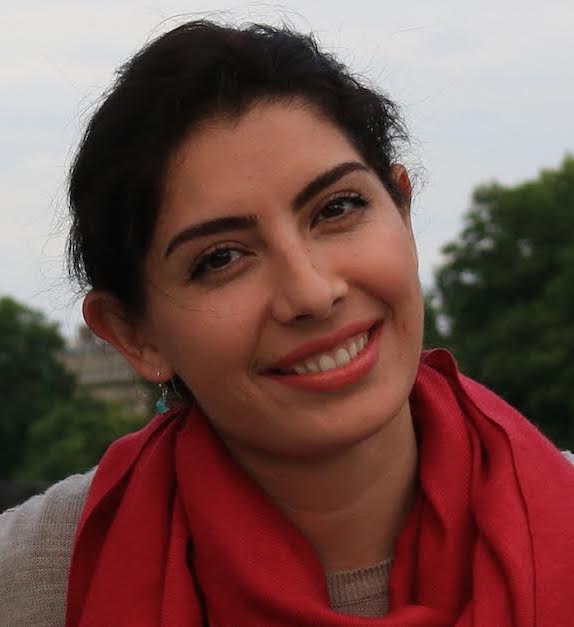}}]
{Elham Naghizade} is a Lecturer at the School of Computing Technologies, RMIT University, Melbourne, Australia. She holds a Ph.D. in Computer Science from the University of Melbourne, where her research focused on data privacy and spatio-temporal data mining. Her research focuses on data privacy, spatio-temporal data mining, explainable time-series classification and responsible AI.
\end{IEEEbiography}

\begin{IEEEbiography}
[{\includegraphics[width=1in, height=1.25in, clip, keepaspectratio]{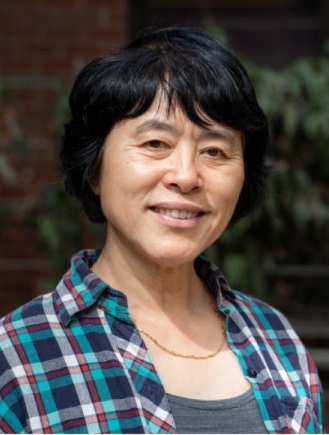}}]
{Fengling Han} (Senior Member, IEEE) received the bachelor’s degree in control theory and application from the Harbin Ship-Building Engineering Institute of Technology, China, the master’s degree in automatic control engineering from the Harbin Institute of Technology, China, and the Ph.D. degree in computer and electronic engineering from RMIT University, Australia. She is Associate Professor at RMIT University. Her current research interests include complex networks, industrial electronics, and cyber security. She has been involved in and leading research projects awarded by the Australia Research Council and the Victoria Government. She is an associate editor and a reviewer for top IEEE journals.

\end{IEEEbiography}

\begin{IEEEbiography}[{\includegraphics[width=1in,height=1.25in,clip,keepaspectratio]{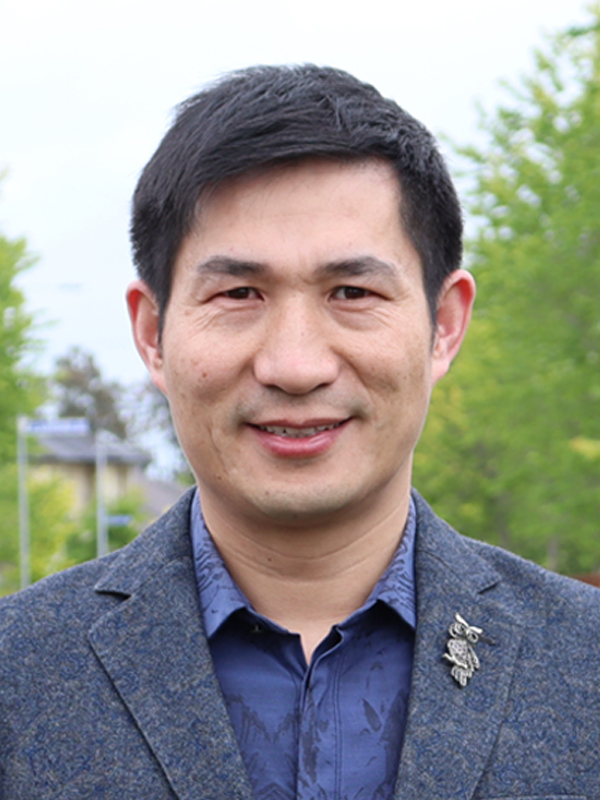}}]{Feng Xia } (Senior Member, IEEE) received the BSc and PhD degrees from Zhejiang University, Hangzhou, China. He is a Professor in School of Computing Technologies, RMIT University, Australia. Dr. Xia has published over 300 scientific papers in journals and conferences (such as IEEE TAI, TKDE, TNNLS, TC, TMC, TBD, TCSS, TNSE, TETCI, TETC, THMS, TVT, TITS, ACM TKDD, TIST, TWEB, TOMM; IJCAI, AAAI, NeurIPS, ICLR, KDD, WWW, MM, SIGIR, EMNLP, and INFOCOM). His research interests include artificial intelligence, graph learning, brain, digital health, and robotics. He is a Senior Member of IEEE and ACM, and an ACM Distinguished Speaker.
\end{IEEEbiography}

\end{document}